\documentstyle[preprint,prd,aps,eqsecnum,epsfig]{revtex}
\begin{document}
%\draft
\title{Extrinsic Curvature Embedding Diagrams}
\author{J. L. Lu ${}^{(1)}$ and W.-M. Suen ${}^{(2,3)}$}
\address{
$^{(1)}$ Physics Department, Hunan Normal University, Hunan, P. R. China
}
\address{$^{(2)}$ 
Department of Physics, Washington University,
One Brookings Drive, St. Louis, MO 63130, USA}
\address{$^{(3)}$ 
Department of Physics, Chinese University of Hong Kong,
Shatin, Hong Kong}
\date{To be submitted to Physical Review D, \today}
\maketitle
\begin{abstract}
Embedding diagrams have been used extensively to visualize the
properties of curved space in Relativity.  We introduce a new kind of
embedding diagram based on the {\it extrinsic} curvature (instead of
the intrinsic curvature).  Such an extrinsic curvature embedding
diagram, when used together with the usual kind of intrinsic curvature
embedding diagram, carries the information of how a surface is {\it
embedded } in the higher dimensional curved space.  Simple examples
are given to illustrate the idea.
\end{abstract}
\newpage
\section{INTRODUCTION}
Embedding diagrams have been used extensively to visualize and understand
properties of hypersurfaces in curved space. They are surfaces in a
fiducial flat space having the same {\it intrinsic} curvature as the
hypersurface being studied.  In this paper we call the former a
``model surface'' and the latter a ``physical surface''.  A familiar
example is the ``wormhole'' construction as the embedding diagram of
the time symmetric hypersurface in the maximally extended
Schwarzschild geometry \cite{MTW}.  Another example often used is a
sheet of paper curled into a cone in the 3 dimensional flat space.
With the intrinsic curvature of the conical surface being zero, the
"model surface" in the embedding diagram is a flat surface.

In this paper we investigate the construction of a different kind of
embedding diagrams.  We examine the construction of a model surface
(in a fiducial flat space) having the same {\it extrinsic} curvature
as the physical surface.  Such an {\it extrinsic} curvature embedding
diagram describes not the geometry of the physical surface, but
instead how it is {\it embedded} in the higher dimensional physical
spacetime.  (For convenient of description, in this paper we will
discuss in terms of a 3 dimensional spacelike hypersurface in the 4
dimensional spacetime.  The same idea applies to a surface of any
dimension in a space of any higher dimensions).

It is of interest to note that such an extrinsic curvature embedding
diagram carries two senses of ``embedding'': (1) It is a surface
``embedded'' in a fiducial flat space to provide a representation of
some properties of the physical surface (the meaning of embedding in
the usual kind of embedding diagram based on intrinsic curvature), and
(2) the diagram is also representing how the physical surface is
``embedded'' in the physical spacetime.  The extrinsic curvature
embedding carries information complimentary to the usual kind of
embedding diagram showing the intrinsic curvature (which we call
``intrinsic curvature embedding'' in this paper).  For example, in the
case of the constant Schwarzschild time hypersurface in a
Schwarzschild spacetime, the {\it extrinsic} curvature embedding is a
flat surface.  For the case of the curled paper, the extrinsic
curvature embedding is a conical surface.

In addition to its pedagogical value (like those of intrinsic
curvature embedding in providing visual understanding), such extrinsic
curvature embedding may help understand the behavior of different time
slicings in numerical relativity, and properties of different
foliations of spacetimes.  Some elementary examples are worked out in
this paper as a first step in understanding extrinsic curvature
embedding.  

\section{Intrinsic v.s. Extrinsic Curvature Embedding}

In the usual kind of embedding diagram (the intrinsic curvature
embedding) one constructs a ``model''
surface in a fiducial flat space which has the same intrinsic geometry
as the physical surface, in the sense of having the same induced
metric.  It should immediately be noted that in general it is
impossible to match all metric components of the two surfaces
\cite{Kasner21c}.  For example, for a 3 dimensional (3D) surface in a
4D curved space, the induced metric $g_{ij}$ ($i,j = 1,2,3$) (the
first fundamental form) has 6 components, each of which is function of
3 variables ($x1, x2, x3$).  The 3D model surface in the fiducial
4D flat space (with flat metric in coordinates ($\tilde{x}^0,
\tilde{x}^1, \tilde{x}^2, \tilde{x}^3$)) is represented by only one
function of 3 variables $\tilde{x}^0 = \tilde{x}^0(\tilde{x}^1,
\tilde{x}^2, \tilde{x}^3)$.  There are 3 more functions one can
choose, which can be regarded either as making a coordinate change in
the physical or model surface, or as choosing the mapping between a
point ($x1, x2, x3$) on the physical surface to a point
($\tilde{x}^1, \tilde{x}^2, \tilde{x}^3$) on the model surface.
Altogether, there are 4 arbitrary functions (e.g., $\tilde{x}^0 =
\tilde{x}^0(\tilde{x}^i), \tilde{x}^i=\tilde{x}^i(x^j), i,j=1,2,3$) at
our disposal.  In general we cannot match all 6 components of the
induced metric.  Only certain components can be matched, and the
embedding can only provide a representation of these components.  An
alternative is to construct an embedding with the model surface
in a higher dimensional space \cite{Fronsdal59,Clarke70,Kasner21b}.

In the case of a stationary spherical symmetric spacetime like the
Schwarzschild spacetime, and when one is examining the geometry of a
constant-Killing-time slice, one can choose a coordinate system (e.g.,
the Schwarzschild coordinate) in which there is only one non-trivial
induced metric component (e.g., the radial metric component).  This component
can be visualized with an embedding diagram using the trivial mapping
$\tilde{x}^1=x1$, $\tilde{x}^2=x2$, $\tilde{x}^3=x3$ between the
physical space and the fiducial space, with $x1 =r$ being the circumferential
radius, $x2 = \theta$ and $x3= \phi$.  This leads to the 
``wormhole'' embedding diagram in textbooks and popular literature.  

Next we turn to extrinsic curvature embedding diagrams.
To illustrate
the idea, we discussed in terms of a 3D spacelike hypersurface in a 4D
spacetime.  
%The same construction can be applied to an
%arbitrary hypersurface in an arbitrary dimensional curved space or spacetime.
Consider a constant time hypersurface in a 4D spacetime with the
metric given in the usual $3+1$ form

\begin{equation}
   dS2 = -(\alpha dt)2 + g_{ij}(dx^i + \beta^i dt)
           (dx^j+\beta^j dt) \quad .
\end{equation}
$\alpha$ is the lapse function, $\beta^i$ is the shift vector, and
$g_{ij}$ is the spatial 3-metric of the constant $t$ hypersurface.
The extrinsic curvature (the second fundamental form) expressed in
terms of the lapse and shift function is
\begin{equation}
  K_{ij} = {1\over 2\alpha} \biggl( \beta_{i/j}+\beta_{j/i} 
           - {\partial g_{ij}\over\partial t} \biggr)\quad .
\label{eq:kij}
\end{equation}
Here ``$/$'' represents covariant derivative in the three-dimensional
space.  

We seek a surface $\tilde{t} = f(\tilde{x}^i)$ with the same extrinsic
curvature $K_{ij}$ embedded in a fiducial 4D flat spacetime
\begin{equation}
   dS2 = -(d \tilde{t})2 + \delta_{ij} d \tilde{x}^i d \tilde{x}^j \quad .
\end{equation}
It is easy to see that
the extrinsic curvature of the surface $\tilde{t} = f(\tilde{x}^i)$ is given by
\begin{equation}
  K_{ij} = {1\over 2 \bar{\alpha}} \biggl( \bar{\beta}_{i/j}+\bar{\beta}_{j/i} 
           \biggr)\quad ,
\label{eq:embedk}
\end{equation}
where $\bar{\alpha} = \sqrt {1+ {\partial f\over\partial
\tilde{x}^i}{\partial f\over\partial \tilde{x}^j} \bar{g^{ij}}}$,
$\bar{\beta}_i = - {\partial f\over\partial \tilde{x}^i}$, and the
covariant derivative in $\bar{\beta}_{i/j}$ is with respect to a
3-metric $ \bar{g_{ij}}$ defined by $ \bar{g_{ij}}= \delta_{ij}-
{\partial f\over\partial \tilde{x}^i}{\partial f\over\partial
\tilde{x}^j}$.  $\bar{g^{ij}}$ is the matrix inverse of $
\bar{g_{ij}}$.

For any given 3 hypersurface in a 4D spacetime, we have only 4
functions that we can freely specify ($f(\tilde{x}^i)$ and the 3
spatial coordinate degrees of freedom), but there are 6 $K^{ij}$
components to be matched.  In general we can only have embedding
representations of 4 of the components of the extrinsic curvature unless
we go to a higher dimensional space,
just like in the case of intrinsic curvature embedding.  This brings 
a set of interesting questions: Under what conditions will a surface be 
fully ``extrinsically-embeddable'' in a fiducial flat space one dimensional higher?  How many dimensions higher must a fiducial space be in order for a general surface to be extrinsically-embeddable?  We hope to return to these questions in future publications.

The two kinds of embedding diagrams, intrinsic curvature embedding and
extrinsic curvature embedding, are supplementary to one another and
can be used together.  The information contained in the usual kind
of intrinsic embedding diagram is partial in the sense that different
slicings of the same spacetime will give different intrinsic curvature
embedding diagrams, and this information of which slicing is used (the
choice of the "time" coordinate) is contained in the extrinsic
curvature embedding.  Similarly, the information given in the
extrinsic curvature embedding is partial, in the sense that the
extrinsic curvature components depend on the choice of the spatial
coordinates, an information that is contained in the intrinsic
curvature embedding.

With the two kinds of embedding diagram constructed together, one can
read out both the induced metric components and the extrinsic
curvature components.  In principle, all geometric properties of the
surface can then be reconstructed, including how the surface is
embedded in the higher dimensional spacetime.  In the following
we give explicit examples of these constructions.
 
\section{EXAMPLES OF EXTRINSIC CURVATURE EMBEDDING DIAGRAMS}

We begin with the simple case of the Schwarzschild metric in Schwarzschild
coordinate,
\begin{equation}
  dS2 = -\biggl( 1-{2m\over r}\biggr)dt2 +\biggl( 1-{2m\over r}\biggr)^{-1} dr2
+r2(d\theta2 + \sin2\theta d\phi2 )\quad .
\label{eq:sch}
\end{equation}
Since the metric is time independent and has zero shift, from
(\ref{eq:kij}) one sees immediately that the constant $t$ slicing has
$K^{ij}=0$ for all i and j.  The ``extrinsic curvature embedding'' is
obtained by identifying a point ($r, \theta, \phi$) to a point
($\tilde{r}, \tilde{\theta}, \tilde{\phi}$) in the fiducial flat space
$dS2 = -d \tilde{t}^2 + d \tilde{r}^2 +\tilde{r}^2(d\tilde{\theta}^2
+ \sin2 \tilde{\theta} d\tilde{\phi}^2 )\quad $, and by requiring the
extrinsic curvatures of the physical surface (embedded in
Schwarzschild spacetime) and the model surface (embedded in flat
spacetime) be the same.  This leads to a flat model surface in the
fiducial flat space.  We see that while the {\it intrinsic} curvature
embedding of the Schwarzschild slicing is non-trivial (as given in
text books and popular articles), the {\it extrinsic} curvature embedding is
trivial.  This high-lights that the constant Schwarzschild time
slicing is a ``natural'' foliation of the Schwarzschild geometry, in
the sense that these (curved) constant-Schwarzschild-$t$ surfaces are
embedded in the (curved) Schwarzschild geometry in a trivial manner:
same as a flat surface embedded in a flat spacetime.

It is interesting to compare this to different time slicings in
Schwarzschild spacetime.  Define 
\begin{equation}
   t = t^\prime 
     + \int { \sqrt{2m\over r}
       \over 1- {2m\over r}} dr \quad .
\end{equation}
The Schwarzschild metric (\ref{eq:sch}) becomes
\begin{equation}
   dS2 = -\biggl( 1- {2m\over r}\biggr) dt^{\prime 2}
          -2\sqrt{{2m\over r}} dt^\prime dr
          + dr2 + r2(d\theta2+\sin2\theta d\phi2 ) \quad .
\label{eq:sch2}
\end{equation}
The $t^\prime={\rm constant}$ surfaces have flat {\it intrinsic}
geometry, so the intrinsic curvature embedding is trivial (the model
surface is a flat surface in the fiducial flat space).  But the
{\it extrinsic} curvature embedding is non-trivial; as we shall work out
below.  This is just the opposite situation of the
constant-Schwarzschild-$t$ slice (non-trivial intrinsic embedding but
trivial extrinsic embedding).

For the extrinsic curvature embedding of the constant-$t^\prime$
``flat slicing'' of metric (3.3), with the spherical symmetry, it
suffices to examine the slice $\theta ={\pi\over 2}$.  A constant
$t^\prime$ slicing in metric (\ref{eq:sch2}) has extrinsic curvature
\begin{eqnarray}
       K_{rr} 
  &=& {1\over 2r} \sqrt{{2m\over r}} \quad , \\
\label{eq:ext1a}
       K_{\phi\phi} 
  &=& - \sqrt{2mr} \quad .  
\label{eq:ext1b} 
\end{eqnarray}
The extrinsic curvature embedding is given
by a $\tilde{t}=f(\tilde{r}, \tilde{\phi})$ surface embedded in a fiducial 3D Minkowski space
\begin{equation}
     dS2 = -d\tilde{t}^2 + d\tilde{r}^2 
            + \tilde{r}^2 d\tilde{\phi}^2 \quad .
\label{eq:flat}
\end{equation}

Using (\ref{eq:embedk}),  it is straightforward to
find that the non-trivial
extrinsic curvature components are
\begin{eqnarray}
       K_{\tilde{r} \tilde{r}} 
   &=& -{f^{\prime\prime} \over{ \sqrt{1-f^{\prime 2}}}} \quad , \\
       K_{\tilde{\phi} \tilde{\phi}}
   &=& -{\tilde{r} f^{\prime} \over\sqrt{1-f^{\prime 2}}} \quad ,
\label{match1}
\end{eqnarray}
where $f^\prime = df/d\tilde{r}$ is a function to be
determined by matching the extrinsic curvature $(K_{\tilde{r}
\tilde{r}} , K_{\tilde{\phi} \tilde{\phi}})$ to that of
the physical surface given by (3.4), (3.5).

%(\ref{eq:ext1a}), (\ref{eq:ext1b}).

It is immediately clear that with only one arbitrary function
$\tilde{t}=f(\tilde{r}, \tilde{\phi})$, it would not be possible to
match both of the two non-trivial extrinsic curvature components.  To
enable the matching, we introduce a spatial coordinate
transformation {\it on} the $t^\prime={\rm constant}$ physical
surface $r = h(r^\prime)$.  As $K_{ij}$ is a tensor on
the surface, the coordinate change will change the value of $K_{rr}$
but {\it not} how the surface is embedded.  Due to the
spherical symmetry, it suffices to rescale only the radial coordinate,
keeping the angular coordinate unchanged.

Using (3.7, 3.8) and (3.4, 3.5), and identifying the
fiducial flat space coordinates $(\tilde{r}, \tilde{\phi})$ with
physical space coordinate $(r^\prime , \phi)$, we obtain the
conditions on the functions $f(\tilde{r})$ and $ h(r^\prime=\tilde{r})$
\begin{eqnarray}
 {({h^\prime})2 \over 2h} \sqrt{{2m\over h}} 
   &=& -{f^{\prime\prime} \over{\sqrt{1-f^{\prime 2}}}} \quad , \\
   \sqrt{2mh} \quad 
   &=& {\tilde{r} f^{\prime} \over\sqrt{1-f^{\prime 2}}} \quad , 
%\label{match1}
\end{eqnarray}
where $h^\prime = dh/d\tilde{r}$.  The boundary conditions for the
system are (i) $f^{\prime}$ tends zero at infinity, and (ii) $h$ tends to
$\tilde{r}$ at infinity; that is, the embedding is trivial
asymptotically.  The two equations lead to a quadratic equation for
$f^{\prime\prime}$ with the two roots
\begin{equation}
   f^{\prime\prime} = - {{f^\prime (1- f^{\prime 2})} \over {4
   \tilde{r}}} \left( (5- f^{\prime 2}) \pm 
 \sqrt{(1- f^{\prime 2})(9- f^{\prime 2})}\right) \quad .
\label{feq}
\end{equation}
While both the ``+'' and the ``-'' sign solutions satisfy the boundary
condition (i) for $f^{\prime}$, it is easy to see that only the ``-''
solution leads to a $h(\tilde{r})$ that satisfies the boundary
condition (ii) for $h$.  Integration of the 2nd order equation associated
with the ``-'' solution gives the extrinsic curvature embedding
diagram for the spatially flat constant time slicing of the
Schwarzschild spacetime as shown in Fig. 1.  The height of the surface
is the value of $f$, the horizontal plane is the $(\tilde{r} , \phi )$
plane (recall $\tilde{r} =r ^{\prime}$).  All quantities are in unit of
$m$ (i.e., $m=1$). 

In what sense does this figure provide a ``visualization'' of the
extrinsic curvature $K_{ij}$ of the physical surface?  The extrinsic
curvature compares the normal of the surface at two neighboring points
(cf. Sec. 21.5 of \cite{MTW}).  In Fig. 1, with the model surface
embedded in a flat space, one can easily visualize (i) unit vectors
normal to the surface , (ii)the parallel transport of a unit normal
vector to a neighboring point, and (iii) the subtraction of the
transported vector from the unit normal vector at the neighboring
point, all in the usual flat space way.  For example, in Fig. 1,
imagine unit normals at two neighboring points $(r, \phi)$ and $(r,
\phi + d \phi)$.  With the horn shape surface, the ``tips'' of the
unit normal vectors are closer than their bases.  When parallel
transported, subtracted and projected into the $\phi$ direction (all
done in the flat space sense) this gives the value of $K_{\phi \phi}$.
On the other hand, if we compare the normals of the neighboring points
$(r, \phi)$ and $(r+dr, \phi)$, the ``tips'' of the normal vectors are
further away than their bases.  This accounts for the difference in
sign of $K_{rr}$ and $K_{\phi \phi}$ in (3.4) and (3.5).  Also
explicit visually is the fact that, at large $r$, the unit normals at
neighboring points (both in the $r$ and $\phi$ directions) become
parallel, showing that the extrinsic curvature goes to zero.  (Notice
that $K_{\phi \phi}$ is not going to zero as $d/d\phi$ is not a unit
vector; rather, the extrinsic curvature contracted with the {\it unit}
vector in the ${\phi}$ direction is going to zero as $r^{-3/2}$ in the
same way as $K_{rr}$.)  We note that $f$ does not tend to a constant
but is proportional to $\sqrt{r}$ at large $r$, although $f^\prime$
does go to zero as implied by the boundary condition.

We note that this prescription of visualizing the covariant components
of the extrinsic curvature $K_{ij}$ is preciously the flat space
version of the prescription given in Sec. 21.5 of \cite{MTW}.  While
the directions of the normal vectors and the result of a parallel
transport are not readily visualizable in the curved space
construction given in \cite{MTW}, the use of an embedding diagram in a
fiducial flat space enables the easy visualization of normal vectors
and their parallel transport--- as all of them are constructed in the
usual flat space sense.  It is also for the easiness of visualization
that we choose to work with the covariant component of the extrinsic
curvature.  While the contravariant components can be treated
equivalently (note that we are working with spacetimes endowed with
metrics), its visualization involved one-form which is less familiar
(see however the visualization of forms in \cite{MTW}).

Returning to the example at hand, we show in Figs. 2a and 2b the
``scaling function'' $h(\tilde{r})$ v.s. $\tilde{r}$.  We see that $h$
is linear in $\tilde{r}$ for large $\tilde{r}$, satisfying the
boundary condition (ii).  In Fig. 2a, we see that $h$ is nearly linear
throughout.  To see that $h$ is not exactly linear, we show in Fig. 2b
that $h^\prime -1$ is appreciably different from zero in the region of
smaller $r$.  This small difference from exact linearity is precisely
what is needed to construct a model surface that can match both
$K_{rr}$ and $K_{\phi \phi}$.

We see that the embedding is perfectly regular at the horizon ($r=2$).
It has a conical structure at $\tilde{r}=0$, 
in the sense that $f^\prime$ is not going to zero but instead
approaches 1 from below (i.e., $f^\prime \sim 1-a2
\tilde{r}^2$ for small $\tilde{r})$.  Although the surface
covers all $\tilde{r}$ values, we note that $h(\tilde{r})$ approaches a
constant $\sim 0.2$, implying that the embedding diagram does {\it
not} cover the inner-most region (from $r=0$ to $r \sim 0.2m$) of the
the circumferential radius $r$.  Comparing this to the
constant-Schwarzschild-time slicing (constant $t$ slicing in metric
(3.1)) is again interesting: The {\it intrinsic} curvature embedding
of the constant-Schwarzschild-time slicing also does not cover the
inner-region (from $r=0$ to $r =2m$), while the {\it
extrinsic} curvature embedding of the constant-Schwarzschild-time
slicing covers all $r$ values just like the {\it intrinsic} curvature
embedding of the ``spatially flat'' slicing.

We emphasize again that the extrinsic curvature embedding diagram
Fig. 1 does {\it not } carry any information about the intrinsic
geometry of the surface.  For example, the circumference of a circle
at a fixed $\tilde{r}$ is not $2 \pi \tilde{r}$, and the distance
on the model surface is not the physical distance between
the corresponding points on the physical surface (unlike
the case of the intrinsic curvature embedding diagram).   This
extrinsic curvature embedding diagram
Fig. 1 carries only the
information of how the ``spatially flat'' slicing is embedded in the
Schwarzschild geometry, in the sense that the relations between the normal
vectors of the slicing embedded in the curved Schwarzschild spacetime
are the same as given by the surface shown in Fig. 1 embedded in a flat
Minkowski spacetime.  

One might want to obtain the physical distance between two neighboring
points, say, at $\tilde{r}$ and $\tilde{r}+ d\tilde{r}$, in Fig. 1.
This information is contained in Figs.2a and 2b, as the scaling factor
$h$ gives the relation between $r$ and $\tilde{r} = r^\prime$.

One can also give a visual representation of this information of the
intrinsic geometry by plotting an {\it intrinsic} embedding diagram,
as in Fig. 2c.  For this spatially flat slicing, the {\it intrinsic}
embedding diagram is a flat surface in a fiducial flat space.  To
enable this intrinsic embedding diagram Fig.2c to be used conveniently
with the extrinsic embedding diagram Fig. 1., we have plotted Fig. 2c
in a way different from what is usually done in plotting embedding
diagrams: The labeling of the spatial coordinate in this diagram is
given in $r^\prime$, the same coordinate (note $\tilde{r}=r^\prime$)
as used in the extrinsic embedding diagram (or more precisely, it is
$x^\prime = r^\prime cos(\phi) $, and $y^\prime = r^\prime sin(\phi)
$).  In this way, the physical distance between any two coordinate
points $r^\prime _1$ and $r^\prime _2$ in the {\it extrinsic}
curvature embedding Fig. 1 (remember $r^\prime = \tilde{r}$, the
coordinate used in Fig. 1) can be obtained directly by measuring the
distance on the model surface between the corresponding two points
$r^\prime _1$ and $r^\prime _2$ in Fig. 2c.  Hence, between this pair
of intrinsic and extrinsic embedding diagrams, we can obtain all
necessary information about the physical surface, with both the first
(metric) and second (extrinsic curvature) fundamental forms explicitly
represented.

We note that in Fig. 2c, the coordinate labels are very close to
equally spaced.  This is a reflection of the fact that the scaling
function $h$ given in Fig. 2a is very close to being linear (but not
exactly).  This near-linearity of the scaling function, together with
the fact the intrinsic embedding diagram is flat, tell us that in this
special case, the physical distances (the physical metric) on the {\it
extrinsic} curvature embedding surface in Fig. 1 between points are,
to a good approximation, given simply by their coordinate separations
in $r^\prime$ (while the extrinsic curvature is contained in the shape
of the surface).  Obviously this would not be true in general.

Next we turn to another simple example.
The infalling Eddington-Finkelstein coordinate $V$ is defined by
$ V \equiv t+r^\star = t+r+2m \ln\left({r\over 2m}-1\right) \quad .$
Let
\begin{equation}
  \bar{t} \equiv V-r=t+2m\ln\left({r\over 2m} -1\right) \quad .
\end{equation}
The Schwarzschild metric in the ``infalling $\bar{t}$ slicing'' becomes
\begin{equation}
   dS2 = -\left( 1-{2m\over r}\right) d\bar{t}^2
          + {4m\over r} d\bar{t}dr +\left(1+{2m\over r}\right)dr2
+ r2(d\theta2+\sin2\theta d\phi2) \quad .
\end{equation}
Both the intrinsic and extrinsic curvature embedding diagrams of the
infalling $\bar{t}$ slicing are non-trivial.  In the following we work
out the extrinsic curvature embedding.

The extrinsic curvature of the ``infalling slicing'' is given by
\begin{equation}
 K_{ij} = \left(
\matrix{ -{2m\over r2} { {1+{m\over r}}\over\sqrt{1+{2m\over r}} }&0&0\cr
         0&{2m\over\sqrt{1+{2m\over r}} }&0\cr
         0&0&{2m\sin2\theta\over\sqrt{1+{2m\over r}} }\cr} \right)\quad .
\end{equation}
Again with the spherical symmetry it suffice to study the slicing $\theta
={\pi\over 2}$.  To construct the extrinsic embedding, we (i) introduce a
coordinate scaling $r=h(r^\prime)$, (ii) identify the coordinate
$(r^\prime, \phi)$ with $(\tilde{r}, \tilde{\phi})$ of (3.6), and (iii)
require $K_{r^\prime r^\prime}=K_{r r} , K_{\phi \phi}=K_{\tilde{\phi}
\tilde{\phi}}$.  This leads to the following equations for $f$ and $h$:
\begin{eqnarray}
       -{ \tilde{r} f^\prime\over\sqrt{1-f^{\prime 2}} }
   &=& { 2m\over\sqrt{1+{2m\over h}} } \quad ,\\
       -{f^{\prime\prime}\over\sqrt{1-f^{\prime 2}} } 
   &=& -{ {2m (h^\prime)2 }\over h2} {1+{m\over h}
         \over \sqrt{1+{2m\over h}} }\quad .
\end{eqnarray}
Eliminating $h$ leads to a quadratic equation for $f^{\prime \prime}$,
the two roots of which give two second order equations for $f$.  We
omit the rather long expressions here.  Again only one of the two
equations admit a solution with the correct asymptotic behavior at
large $\tilde{r}$ ($f^{\prime}$ tends to zero and $h$ tends to
$\tilde{r}$).  Integrating this second order equation gives the
embedding diagram shown in Fig. 3.  The height of the surface
represents the value of $f$, the horizontal plane is the ($\tilde{r},
\phi$) plane.  All quantities are in unit of $m$.  Fig. 4a gives the
scaling function $h(\tilde{r})$ v.s. $\tilde{r}$, showing that it
satisfies the boundary condition at infinity.  Asymptotically $h$
tends to $\tilde{r}$, while $f \sim -2mlog(\tilde{r})$, and
$f^{\prime} \sim {-2m\over r} + {{2m2}\over {r2}}$.  Again we see
that $h$ is very close to being linear.  To show that it is not
exactly linear, we plot in Fig. 4b the derivative of $h$
v.s. $r^\prime$.  For $r^\prime <2m$, the derivative is considerably
less than $1$.  

As one may expect, the embedding is regular at the horizon, but
has a conical structure at $\tilde{r}=0$, same as the ``spatially flat
slicing'' case above.  For small $\tilde{r}$, $f^\prime$ tends to
$-1$ (from above), while $h$ tends to a constant $\sim 1.2m$.  This
implies that the inner most region of the circumferential radius $r$
(from 0 to 1.2m) is not covered in the embedding diagram, again similar to
the ``spatially flat slicing'' extrinsic curvature
embedding studied above.  

We see that while the model surface in the ``spatially flat slicing''
embedding diagram Fig. 1 dips down for small $\tilde{r}$, the model
surface in the ``infalling slicing'' embedding diagram Fig. 3 spikes
up.  This is expected as the signs of the extrinsic curvature
components ($K_{rr}, K_{\phi \phi}$) are opposite of one another for
the two slicings.  We can easily see in Figs. 1 and 3, that in one
case ``the tips of the normal are closer than their base'' or vise
versa.  Such visual inspection is possible as the model surfaces are
now embedded in flat spaces, enabling the use of flat space measure
of distances, and normal vectors. 

Again, one might want to visualize the physical distance between two
neighboring points in Fig. 3.  This can be done by plotting the
corresponding {\it intrinsic} embedding diagram in the $r^\prime$
coordinate, as is given in Fig. 4c.  The physical distance between any
two coordinate points $r^\prime _1$ and $r^\prime _2$ can be measured
by their distance on this {\it intrinsic} embedding surface, in the
flat space way.  Due to the near linearity of the scaling function
$h$, we see that the coordinate labels are again very close to equally
spaced.  However, in this case, unlike the spatially flat slicing
above, the physical distance between the same coordinate distance
$dr^\prime$ is larger for smaller $r^\prime$, as we can see from the
curving of the intrinsic embedding surface.  Between this pair of
intrinsic and extrinsic embedding diagrams, we can again visualize all
information of the physical surface.

\section{Summary And Discussions}

In this paper we propose a new type of embedding diagram, i.e., the
``extrinsic curvature embedding diagram'' based on the 2nd fundamental
form of a surface.  It shows how a surface is embedded in a higher
dimensional curved space.  It carries information complimentary to the
usual kind of ``intrinsic curvature embedding diagram'' based on the
1st fundamental form of the surface.  We illustrate the idea with 3
different slicings of the Schwarzschild spacetime, namely the constant
Schwarzschild $t$ slicing (Eq. (3.1)), the ``spatially flat'' slicing (Eq. (3.3)) and the
``infalling'' slicing (Eq. (3.13)).  The intrinsic and extrinsic curvature
embeddings of the different slicings are discussed, making interesting
comparisons.

The intrinsic curvature embedding diagram depends on the choice of the
``time'' slice (in the 3+1 language of this paper), which is a piece
of information carried in the extrinsic curvature.  On the other hand,
the extrinsic curvature embedding diagram constructed out of the
extrinsic curvature components depends on the choice of the
``spatial'' coordinates, which is a piece of information carried in the
intrinsic curvature embedding diagram.  With the two kinds of
embedding diagram constructed together, all geometric properties of
the surface can then be reconstructed, including how the surface is
embedded in the higher dimensional spacetime.

Why do we study embedding diagrams?  One can ask this questions for both the intrinsic and extrinsic embedding constructions.  It is clear that embedding construction has pedagogical value, e.g., the wormhole diagram of the Schwarzschild geometry appears in many textbooks introducing the ideas of curved spacetimes.  The usual embedding diagrams shown are those based on the intrinsic curvature.  Here we introduce a complimentary kind of embedding diagrams which is needed to give the full information of the surface in the curved spacetime. 
Beyond their pedagogical value, we would like to point out that embedding diagram could be useful in numerical relativity.  Indeed the authors were led to the idea of extrinsic curvature embedding in trying to find a suitable foliation (to choose the lapse function) in the numerical construction of a black hole spacetime.  In the standard 3+1 formulation of numerical relativity, the spatial metric $g_{ij}$ and the extrinsic curvature $k_{ij}$ are used in parallel as the fundamental variables in describing a particular time slice.  One chooses a lapse function to march forward in time.  A suitable choose is crucial to make both the $g_{ij}$ and $k_{ij}$ regular, smooth and evolving in a stable manner  throughout the spacetime covered by the numerical construction.  Whether a choice is suitable depends on the properties of the slicing and hence has to be dynamical in nature.  This is a problem not fully resolved even in the construction of a simple Schwarzschild spacetime.  Embedding diagrams let us see the pathology of the time slicing clearly and hence could help in the picking of a suitable lapse function.  For example, in the constant Schwarzschild time slicing (Eq.(3.1)), the intrinsic curvature embedding dips down to infinity at $r=2m$ and cannot cover the region inside (the extrinsic curvature embedding is flat and nice for all $r$).  In the time slicing of Eq. (3.3), the intrinsic curvature embedding is flat and nice for all $r$, but the extrinsic curvature embedding has a conical singularity near $r=0.2m$ and cannot cover the region inside, as shown in Sec. 3 of this paper.  For the use of embedding diagrams in numerical relativity, and in particular in looking at the stability of numerical constructions with different choices of time slicing, one would need to investigate the two kinds of embedding diagrams in dynamical spactimes.
We are working on simple cases of this presently.  

\section{Acknowledgment}
We thank Malcolm Tobias for help in preparing the figures.
This work is supported in part by US NSF grant
Phy 9979985.

\newpage     
\begin{figure}
\caption{Embedding diagram for the ``spatially flat slicing'' of the
Schwarzschild spacetime (line element (3.3)).  The function $f$ given
by (3.11) is plotted on the $\theta = \pi /2$ plane.  $df/d
\tilde{r}$ tends to 1 at the origin ($ \tilde{r}$ tends to 0), where the
embedding has a conical structure.  All
quantities are in unit of m. }
\end {figure}

%\begin{figure}
%\caption{
Fig. 2a. Scaling factor $h$ defined by (3.9, 3.10) for the ``spatially
flat slicing'' of the Schwarzschild spacetime (line element
(3.3)). $h$ tends to $\tilde{r}$ at infinity and is basically linear
through out.  It tends to a non-zero constant $\sim 0.2m$ as
$\tilde{r}$ approaches zero.  

Fig. 2b. Derivative of $h$ with respect to
$\tilde{r}$ is plotted in the close zone, showing that it is not
exactly linear.  This slight deviation from exact nonlinearity is
needed to enable both $k_{rr}$ and $k_{\phi \phi}$ be matched.

Fig. 2c. The {\it intrinsic} curvature embedding diagram
(corresponding to the {\it extrinsic} curvature embedding diagram in
Fig.1) is plotted in $r^\prime=\tilde{r}$, the same coordinate as used
in Fig. 1 (or more precisely, it is $x^\prime = r^\prime cos(\phi) $,
and $y^\prime = r^\prime sin(\phi) $).  The physical distance between
any two coordinate points $r^\prime _1$ and $r^\prime _2$ in the {\it
extrinsic} curvature embedding Fig. 1 can be obtained directly by
measuring the distance in the flat space sense between the
corresponding two points $r^\prime _1$ and $r^\prime _2$ on the model
surface in Fig. 2c.  Between Fig. 1 and 2c, we can obtain all
necessary information about the physical surface, with both the first
(metric) and second (extrinsic curvature) fundamental forms explicitly
represented.

%\end {figure}
%\begin{figure}
%\caption{

Fig. 3. Embedding diagram for the ``infalling slicing'' of the
Schwarzschild spacetime (line element (3.13)).  The function $f$
defined by (3.15, 3.16) is plotted on the $\theta = \pi /2$ plane.
$df/d \tilde{r}$ tends to 1
at the origin ($ \tilde{r}$ tends to 0), where the
embedding has a conical singularity.  All quantities are in unit
of m.

%\end {figure}
%\begin{figure}
%\caption{

Fig. 4a. Scaling factor $h$ defined by (3.15, 3.16) for the ``spatially flat
slicing'' of the Schwarzschild spacetime (line element (3.13)). $h$
tends to $\tilde{r}$ at infinity and is nearly linear through out.
It tends to a non-zero constant $\sim 1.2m$ as $\tilde{r}$ approaches
zero. 

Fig. 4b.  Derivative of $h$ with respect to $\tilde{r}$ is plotted in
the close zone for the infalling slicing, showing that it is not
exactly linear.

Fig. 4c. The {\it intrinsic} curvature embedding diagram for the
infalling slicing, corresponding to the {\it extrinsic} curvature
embedding diagram in Fig.3, is plotted in $r^\prime =\tilde{r}$, the
same coordinate as used in Fig. 3.  Due to the linearity of $h$ in
Fig. 4a, the coordinate labels are nearly equally spaced.  The physical
distance between any two coordinate points $r^\prime _1$ and $r^\prime _2$ 
in the {\it extrinsic} curvature embedding Fig. 3 can be obtained
directly by measuring the distance in the flat space sense on the
model surface Fig. 4c between the corresponding two coordinate points
$r^\prime _1$ and $r^\prime _2$.  We see that the same coordinate
separation corresponds to a large physical distance in the near zone.

%\end {figure}
%\newpage
\begin{figure}
\hspace{-4.1cm}
\vspace{-3.2cm}
\psfig{figure=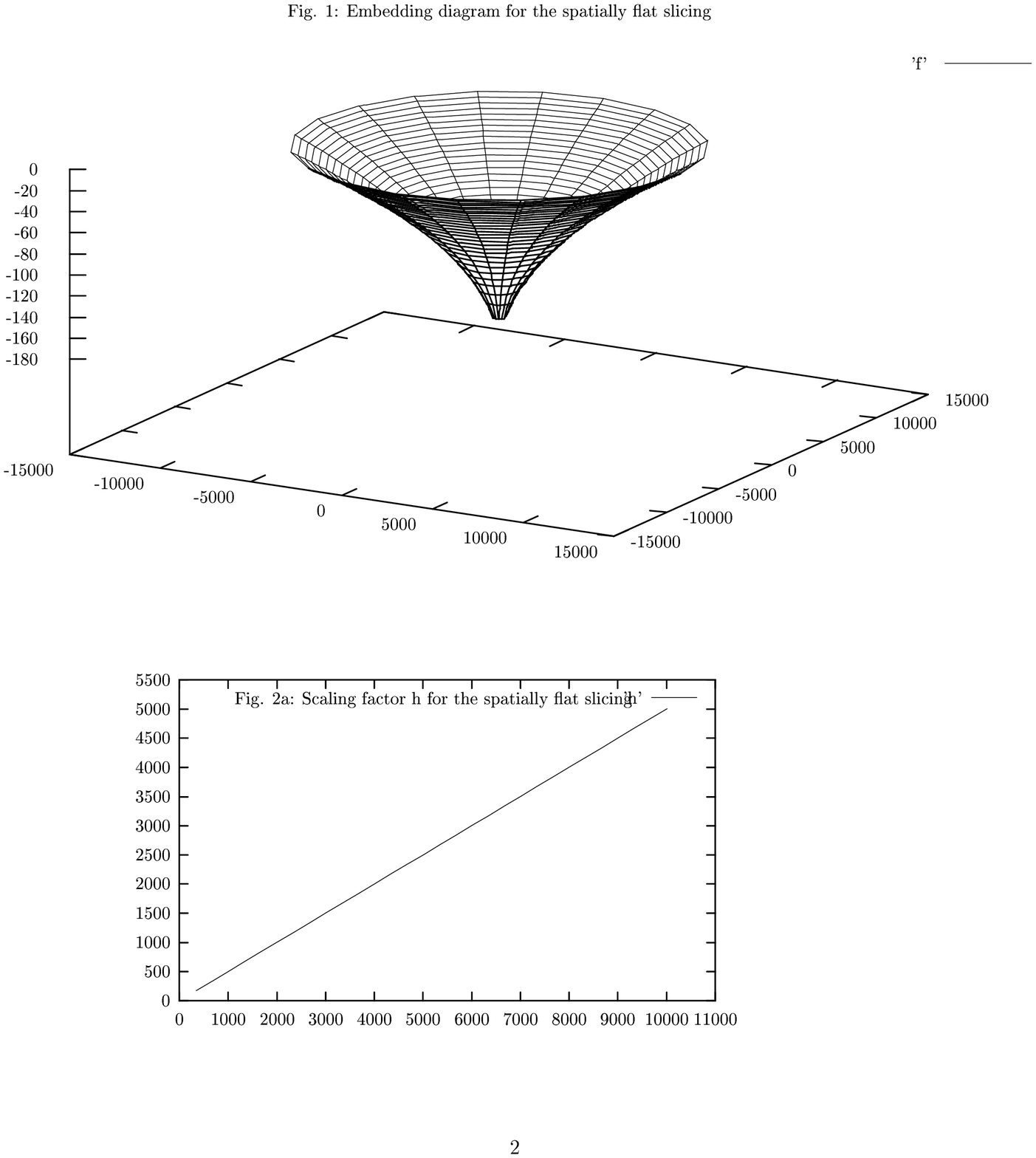,width=20.0cm}
\end{figure}
\begin{figure}
\hspace{-4.1cm}
\vspace{-3.2cm}
\psfig{figure=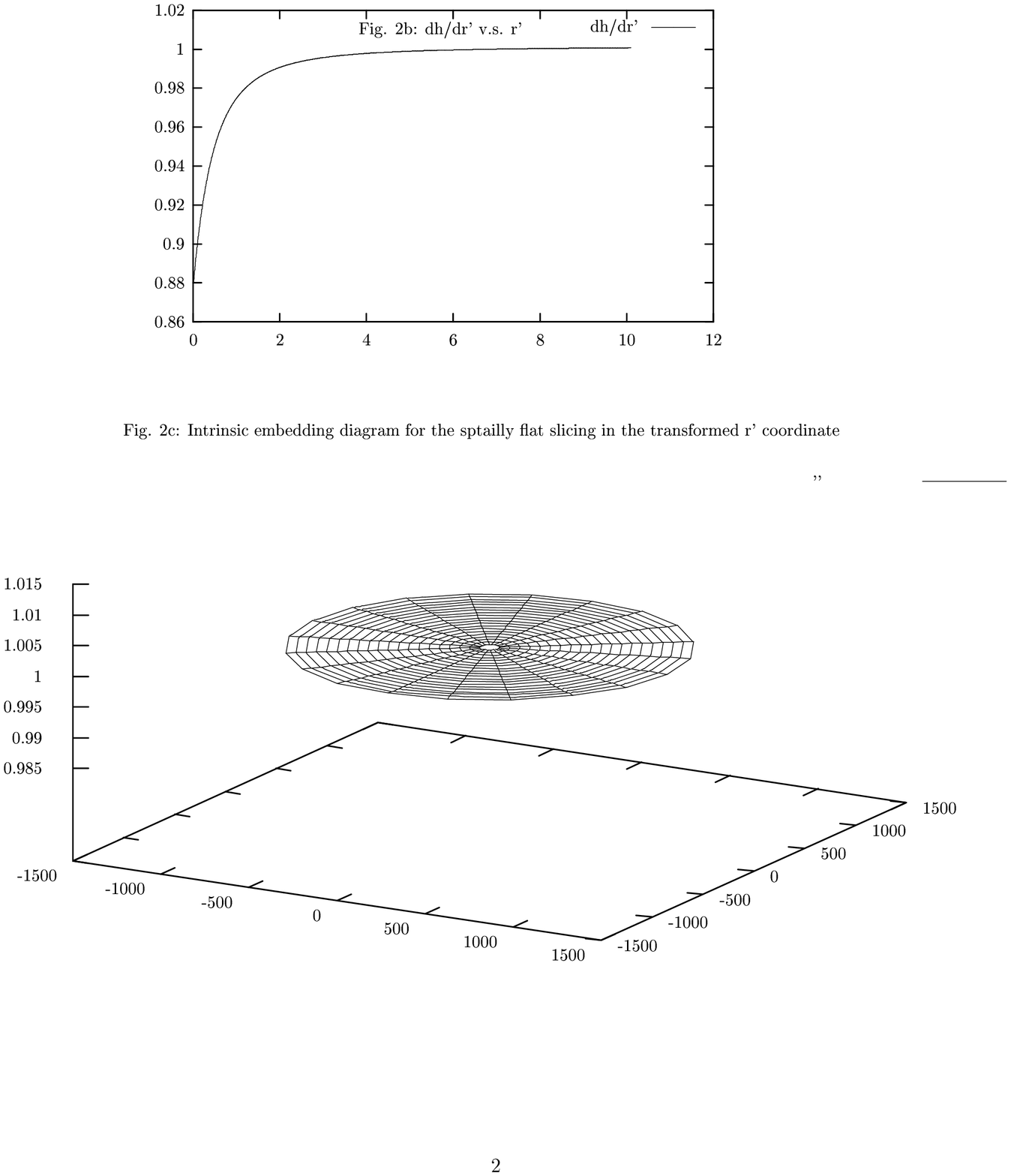,width=20.0cm}
\end{figure}
\begin{figure}
\hspace{-4.1cm}
\vspace{-3.2cm}
\psfig{figure=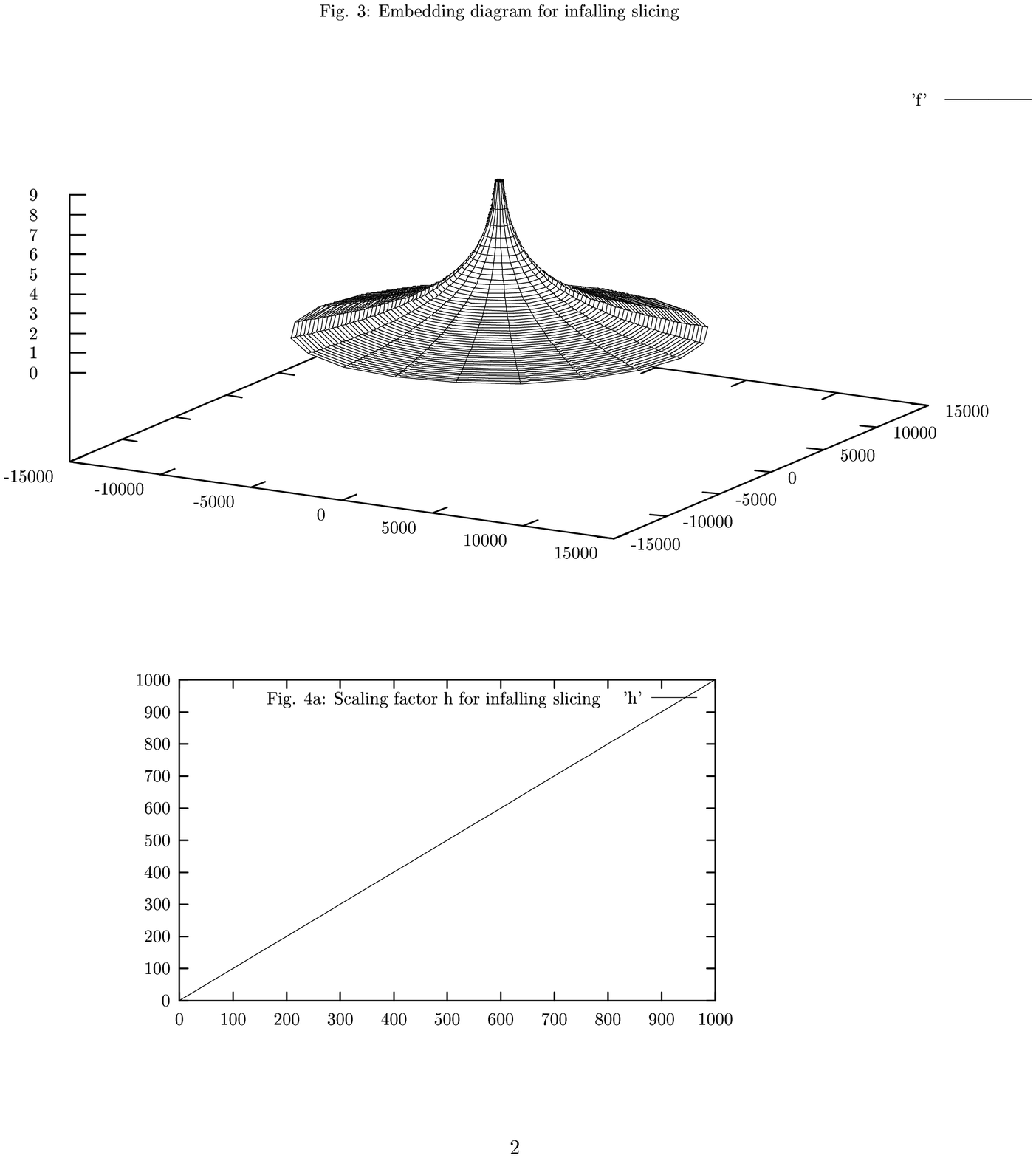,width=20.0cm}
\end{figure}
\begin{figure}
\hspace{-4.1cm}
\vspace{-3.2cm}
\psfig{figure=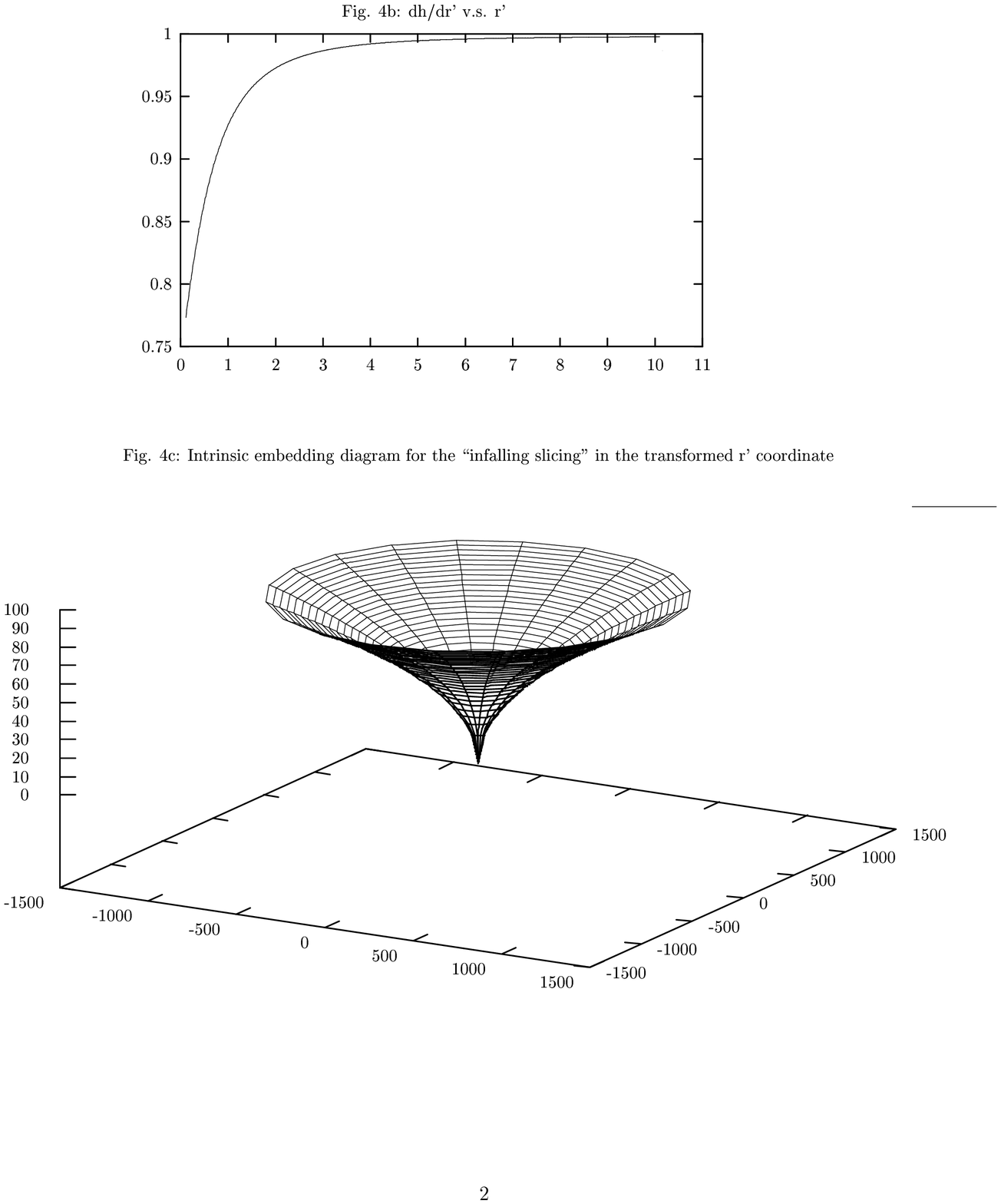,width=20.0cm}
\end{figure}
\end{document}